\documentclass[twoside,leqno,twocolumn]{article}  
\usepackage{ltexpprt}

\usepackage{epsfig}
\usepackage{epstopdf} 
\usepackage{url}
\usepackage{algorithm}
\usepackage{algorithmic,subfigure}
 \usepackage{amsbsy,amsmath, amssymb,comment,nccmath}
\usepackage{color}\usepackage{hhline}\usepackage{colortbl}
\usepackage{url}
\definecolor{myblue}{rgb}{0.6,0.8,1}
\definecolor{mygray}{rgb}{0.93,0.93,0.93}

\newcommand{\hide}[1]{}\newcommand{\m}{\mathcal}\newcommand{\mf}{\mathfrak} 

\newcommand{\bth}{\begin{theorem}}\newcommand{\ethe}{\end{theorem}}
\newcommand{\bpr}{\begin{proof}}\newcommand{\epr}{\end{proof}}
\newcommand{\ble}{\begin{lemma}}\newcommand{\ele}{\end{lemma}}
\newcommand{\bco}{\begin{corollary}}\newcommand{\eco}{\end{corollary}}
\newtheorem{definition}{Definition}

\begin{document}
\date{}

\title{On Randomly Projected Hierarchical Clustering \\ with Guarantees}

\author{Johannes Schneider\\ IBM Research, Zurich, Switzerland \and Michail Vlachos\\ IBM Research, Zurich, Switzerland \thanks{The research has received funding from the European Research Council under the European Union's 7th Framework Programme (FP7/2007-2013) / ERC grant agreement $n^{o}$ 259569.}} 


\maketitle
\begin{abstract}\footnote{This version contains additional details, ommitted in the conference paper ``On Randomly Projected Hierarchical Clustering with Guarantees'', SIAM International Conference on Data Mining (SDM), 2014.}
Hierarchical clustering (HC) algorithms are generally limited to small data instances due to their runtime costs. Here we mitigate this shortcoming and explore fast HC algorithms based on random projections for single (SLC) and average (ALC) linkage clustering as well as for the minimum spanning tree problem (MST).
We present a thorough adaptive analysis of our algorithms that improve prior work from $O(N^2)$ by up to a factor of $N/(\log N)^2$ for a dataset of $N$ points in Euclidean space. 
The algorithms maintain, with arbitrary high probability, the outcome of hierarchical clustering as well as the worst-case running-time guarantees. We also present parameter-free instances of our algorithms. 
\end{abstract}

\section{Introduction}\label{sec:intro}
Despite the proliferation of clustering algorithms (K-means-based, spectral, density-based, statistical), hierarchical clustering (HC) algorithms still are one of the most commonly used variants.
This is attributed to their parameterless nature and their simplicity of implementation.  
No cluster parameter is required because HC builds a nested hierarchy of the clusters formed, often visualized using dendrograms. Dendrograms can be `cut' at any level to form the desired number of clusters. 


Many HC algorithms run in $\Omega(N^2)$ time because all pairwise distances are computed. 
This limits their applicability to rather small datasets. Solutions exist that attempt to mitigate the large runtime by means of approximations, heuristic search, projections, etc. 
As we will discuss in depth in the Related Work Section, 
many of these algorithms come with several assumptions or traits that impose certain limitations: 
complex parameter choices, applicability for specific linkage functions (e.g. only for single-linkage clustering),
or lack of error bounds on the approximation quality.

In this work, we investigate how random projections can be combined with hierarchical
clustering to offer expedient construction of dendrograms with \textit{provable} quality guarantees.
We exploit the fact that for a random projection of high-dimensional points onto a line, 
close points in the high-dimensional space are expected to remain closer than points that
 are far away from each other. This inherent partitioning into groups of similar points forms 
 the basis of our clustering algorithms. Our algorithms leave a small probability for failure. 
 By performing several projections, this probability can be made arbitrarily small. 

Our work contributes to improving the scalability of classical clustering algorithms, such as 
single-linkage (SLC) and average-linkage (ALC) hierarchical clustering,
while offering strict quality guarantees. The key contributions are as follows:
\begin{itemize} 
\item The first algorithm for ALC with $O(dNB\log^2 N)$ running time for a data-dependent number $B$. Both the correctness and running time are guaranteed with probability $1-1/N^c$ for an arbitrarily large constant $c$. This yields a speedup of up to $N/\log^2 N$ compared with prior work.  

\item The first parameter-free algorithms computing a MST, SLC and ALC running in $O(dN polylog(N))$ time for a large class of data. Both the correctness and running time are guaranteed with probability $1-1/N^c$ for an arbitrarily large constant $c$.

\item Introduction of a single data property $B$ for adaptive analysis to capture the computational difficulty of hierarchical clustering.
\end{itemize}

\subsection{Overview} \label{sec:over}
Our clustering algorithms consist of two phases. The first phase is the same for all algorithms and partitions the data into sets of close points. The algorithm PerturbMultiPartition, described in Section \ref{sec:Part}, constructs sets of neighboring points by first perturbing the points given to deal with unfavorable point arrangements and then partitioning them using random projections onto randomly chosen lines. The distances between points on a projected line are scaled by the same factor for all points in expectation. So, close points in the high-dimensional space remain close on the random line and distant points remain distant.  The second phase uses the sets resulting from the partitioning to create clusters efficiently by computing distances only for pairs of points contained in the same set.
Thus, instead of computing the entire distance matrix of size $\Omega(N^2)$ for $N$ points, our algorithm generally requires much fewer distance computations. Specifically, the number of required distance computations depends on the density of data points as discussed and formalized in Section \ref{sec:prop}. 





\section{Related Work} \label{sec:previous}
The (asymptotically) fastest algorithms for SLC and ALC take O($N^2$) \cite{Sib73,Mur84}. For a wide family of metrics, O($N^2\log N$) time algorithms exist \cite{Day84}. In \cite{Kog07} Locality Sensitive Hashing (LSH) was used to speed up SLC. The algorithm runs in $O(N\cdot bucketsize)$ time. However, no guarantees are provided on the clustering quality and, in addition, the algorithm requires the setting of several (non-trivial) parameters, such as, the bucket size and the number of hash functions. 

In \cite{Gio00} LSH is used for implementing a heuristic for complete-linkage clustering. The algorithm 
also requires the setting of several parameters and no guarantees regarding the performance or quality of the clustering are given.
For Euclidean spaces, several MST (SLC) algorithms have been developed for low dimensions, e.g. \cite{Agr90}. 
In \cite{Wil10} a dual-tree algorithm for Euclidean spaces is given. From a traditional worst-case perspective the algorithm behaves worse than prior work, i.e., \cite{Agr90}. However, using adaptive analysis, i.e., introducing the problem-dependent parameters $c,c_p,c_l$, the run time is shown to be $O(\max\{c^6,c_p^2c_l^2\}c^{10}\cdot N \log N \alpha(N))$, where $\alpha(N)$ is a slowly growing function in $N$. The parameters $c,c_p,c_l$ cover expansion properties. They can have any value $[1,N]$. 



In \cite{Urr07} random projections are used for speeding up density-based clustering, where $N$ points are projected onto $n$ random lines. The algorithm runs in time $O(n^2\cdot N + N\cdot \log N)$ and requires several tuning parameters. In \cite{Fern03}, high-dimensional points are projected to a random five-dimensional space, on which an expectation maximization algorithm is run. No analysis is given.

The running time of our algorithm depends on the maximal number of points $|B(P,r)|$ within some distance of a point $P$ relative to the overall number of points $N$. Expander graphs, which have many applications in computer science and beyond \cite{Mar10}, are good examples in which the size of a neighborhood (typically) grows constantly with distance and our algorithms performs well.

\section{Preliminaries} \label{sec:pre}

\begin{figure}[!ht]
\centerline{\includegraphics[width=0.7\linewidth]{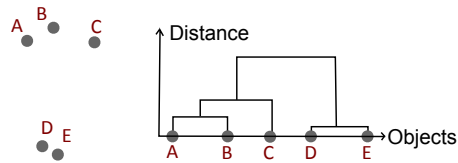}}
\caption{Five points and the corresponding dendrogram according to the single linkage criterion.} 
\label{fig:HCCluster}
\end{figure}

To construct a hierarchical clustering, objects are iteratively merged based on their distance
until only one object group remains. The various HC algorithms basically differ on the distance definition when merging groups of objects. Single-Linkage clustering (SLC) considers the closest distance between any of the objects in the groups as the distance between those groups. For an example, see Figure \ref{fig:HCCluster}. When all distances between objects are averaged, we are talking about Average-Linkage clustering (ALC). Interestingly, the Minimum Spanning Tree (MST)
and the SLC are closely related, as the process involved is the same. To compute an MST, one can iteratively pick the smallest edge $e$ and add it to the tree such that no cycles occur. So the sequence for the MST problem also yields a SLC. 

We use the term \emph{whp}, i.e., with high probability, to denote probability $1-1/N^c$ for an arbitrarily large constant $c$. The constant $c$ (generally) also occurs as a factor hidden in the big $O$-notation. 
To avoid dealing with special cases, all points are assumed to have distinct distances, i.e., $\forall A,B,P,Q \in \m{P}$ holds $D(A,B)\neq D(P,Q)$.

We often use the following Chernoff bound:

\bth \label{thm:Che}
The probability that the number $X$ of occurred independent events $X_i \in \{0,1\}$, i.e. $X:=\sum_i X_i$, is not in $[(1-c_0)\mathbb{E}[X],(1+c_1)\mathbb{E}[X]]$ with $c_0\in]0,1]$ and $c_1\in]0,1[$ can be bounded by $p(X \leq (1-c_0)\mathbb{E}[X] \vee  X \geq (1+c_1)\mathbb{E}[X]) < 2e^{-\mathbb{E}[X]\cdot \min(c_0,c_1)^2/3}$
\ethe
If an event occurs whp for a point (or edge) it occurs for all whp. The proof uses a standard union bound.

\medskip
\bth \label{thm:depEv}
For $n^{c_0}$ (dependent) events $E_i$ with $i \in [0,n^{c_0}-1]$ and constant $c_0$ s.t. each event $E_i$ occurs with probability $p(E_i)\geq 1- 1/n^{c_1}$ for $c_1 > c_0+2$, the probability that all events occur is at least $1-1/n^{c-c_0-2}$.
\ethe

By $c_0(c_1)$ we denote a constant $c_0$ such that $c_0$ tends to 1 if constant $c_1$ tends to infinity, i.e., $c_0\xrightarrow{c_1 \rightarrow \infty} 1$.


\section{Partitioning of Data} \label{sec:Part}
\begin{figure}[!ht]
\centerline{\includegraphics[width=.9\linewidth]{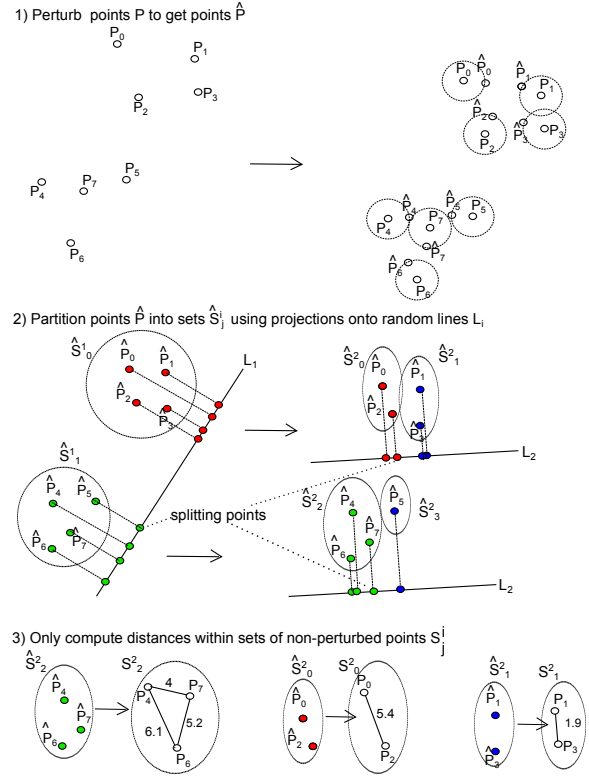}}

\caption{The points $\m{P}$ are first perturbed to yield points $\hat{\m{P}}$. All points $\hat{\m{P}}$ are projected onto line $L_1$ and split into two sets $\hat{\m{S}}^{1}_0,\hat{\m{S}}^{1}_1 \subseteq \hat{\m{P}}$ at a random projected ``splitting'' point. The original perturbed (not the projected) points in set $\hat{\m{S}}^{1}_0 \subseteq \hat{\m{P}}$ are then projected onto $L_2$ in the same manner to yield sets $\hat{\m{S}}^{2}_0,\hat{\m{S}}^{2}_1$, and so on. After the partitioning, the HC algorithm only computes distances among points within sets of non-perturbed points.} 
\label{fig:basicpro}
\end{figure}

An overview of the whole partitioning process is shown in Figure \ref{fig:basicpro}.

\begin{table}[t]
\caption{Notation used in the paper} \label{table:Symb}
\centering 
{\footnotesize
\begin{tabular}{l | l } 
\hline\hline 
Symbol & Meaning \\ 
\hline 
$P,T,Q$ & points in Euclidean space $\mathbb{R}^d$ \\
$\m{P},\m{S},\m{C}$ & set of points \\
$L$ & randomly chosen line \\
$\m{L}$ & sequence of lines $(L_0,L_1,...)$\\
$\mf{S},\mf{W}; \mf{L}$ & set of sets of points; set of sequences of lines\\
$D(A,B)$ & distance, ie. $L2$ norm $||B-A||_2$, for points $A,B$\\
$\m{C}(P)$ & cluster, i.e., set of points, containing point $P$ \\
$ID(P)$ & (arbitrary) unique ID of point $P$ \\
$\m{C}_{ID}(Q)$ & cluster ID: arbitrary ID of a point $P \in \m{C}(Q)$ \\ 
$e=\{P,T\}$ &  undirected edge $e$ between points $P$ and $T$ \\
$\m{HC}$ & sequences of edges $(e_0,e_1,e_2,...,e_{N-2})$ where \\ & $e_i$ caused the $i$th merger of two clusters\\
$\m{HC}(P)$ & all edges $e \in \m{HC}$ with $P$ adjacent to $e$, i.e., $P \in e$ \\
$MST$ & minimum spanning tree consisting of $N-1$ edges \\
$N(P)$ & neighbors of $P$ in the HC (or MST)\\ & i.e.$\{Q|\exists e=\{P,Q\} \in \m{HC} \}$
\end{tabular}
}
\end{table}

Our goal is to obtain small sets of points which are close to each other. When doing a single random projection, in expectation, nearby points remain nearby and distant points remain distant. Therefore, if we split the points projected onto the line into two sets (see Figure \ref{fig:basicpro}), we are more likely to separate pairs of distant points than pairs of close points. By repeating the splitting procedure recursively, the odds are good to end up with small sets of close points. Using multiple partitionings allows us to guarantee, with high probability, that for a fraction of all sets, each pair of nearby points will be contained in at least one of the small sets.

More precisely, we begin by splitting the point set into smaller sets, which are used for clustering (Partition algorithm). We create multiple of these partitions by using different random projections and perturbed points (PerturbMultiPartition algorithm). Intuitively, if the projections $P\cdot L$ and $Q\cdot L$ of two points $P,Q$ onto line $L$ are of similar value then the points should be close. Thus, they are likely kept together whenever the point set is divided.

\begin{figure}[htp]
\centerline{\includegraphics[width=0.95\linewidth]{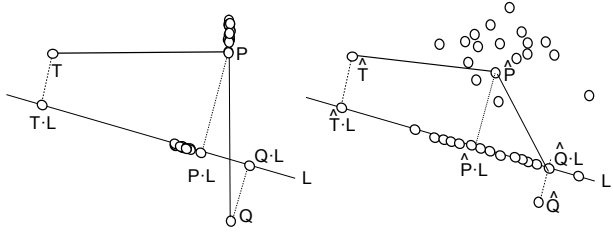}}

\caption{In the left panel it is probable that all points near $P$ are projected in between edge $\{P,T\}$, but none is projected in between the longer edge $\{P,Q\}$. If all points (in particular $P,T,Q$) are perturbed (right panel) then likely at least a constant fraction of all the points are projected in between any adjacent edge $\{P,Q\}$.}
\label{fig:densityperturb}
\end{figure}

For a single partition, we start with the entire point set and add noise to the points to deal with worst-case point arrangements, as shown in Figure \ref{fig:densityperturb}. Then the point set is split recursively into two parts until the size of the point set is below a certain threshold, less than $minPts$. To split the points, the points are projected onto a random line, and one of the projected points on that line is chosen uniformly at random. \footnote{The random line is given by a vector originating at 0 to a randomly chosen point on the $d$-dimensional unit sphere, e.g. using one of the techniques from \cite{Ran12}.\label{foot:rand}}  All points with a projected value smaller than that of the chosen point constitute one part and the remainder the other part. Two consecutive partitionings using random lines $L_1$ and $L_2$ are illustrated in Figure \ref{fig:basicpro}.

A key point is that our algorithms only consider edges $\{P,Q\}$ for merging and distance computations that end up in the same set of size at most $minPts$ after a partitioning. For an example, see Figure \ref{fig:defs} which shows three sets and all corresponding edges taken into account by our algorithms. 

\begin{algorithm}[h!]
{\footnotesize
\caption{PerturbMultiPartition(points $\m{P}$, $minPts$, $l_{per.}$) return set of point sets $\mf{S}$ for $i\in[0,\log N-1]$  \label{alg:PerturbMulti}}
\begin{algorithmic}[1]
\begin{small}
\STATE Choose sequences of random lines $\m{L}^i:=(L_0,L_1,...,L_{c_1\log N})$ for $i\in[0,c_0\cdot \log N-1]$ for constants $c_0,c_1$ with $L_j \in \mathbb{R}^d$ being a random vector of unit length
\FOR{$i=1..c_0\cdot \log N$}
  \STATE $\m{\hat{P}}:=\{P+RV|P \in \m{P}, RV = $ random vector of length  $l_{per.}$ chosen separately for each  $P\}$ \COMMENT{Perturb points}
  \STATE $\mf{\hat{W}}:=$ result of $Partition(\m{\hat{P}},0,i)$ 
  \STATE $\mf{W}:= \{ \{P \in \m{P}| \hat{P}:=P+RV \in \m{\hat{W}}\}| \m{\hat{W}} \in \mf{\hat{W}} \}$ 
  \STATE $\mf{S} := \mf{S} \cup \mf{W}  $
\ENDFOR
\medskip
\medskip 
\\{\textbf{Algorithm Partition}(points $\hat{\m{S}}$, line $j$, sequence $i$) return  set of sets $\mf{\hat{S}}'$} 
\smallskip
\IF{$|\hat{S}|\geq minPts$}
	\STATE Choose $\hat{P} \in \hat{\m{S}}$ uniformly at random
	\STATE $\hat{\m{S}}_0 := \{Q \in \hat{\m{S}}| Q\cdot L_j \leq \hat{P}\cdot L_j, L_j \in \m{L}^i\}$
	\STATE $\hat{\m{S}}_1 :=\hat{\m{S}}\setminus \hat{\m{S}}_0$
	\STATE $Partition(\hat{\m{S}}_0,j+1,i)$
	\STATE $Partition(\hat{\m{S}}_1,j+1,i)$
\ELSE
\STATE $\mf{\hat{S}}':=\mf{\hat{S}}' \cup \{\hat{\m{S}}\}$
\ENDIF
\end{small}

\end{algorithmic}
\small
\smallskip
} 
\end{algorithm}

\medskip
\bth\label{thm:Part}
Algorithm Partition uses O($\log N$) projection lines. It runs in $O(d N \log N)$ time whp.
\ethe
As each projection partitions a point set into two non-empty point sets until a set is below the threshold size $minPts$, it becomes clear that the algorithm eventually terminates. In the proof we use the fact that with constant probability a split creates two sets of a size that is only a constant fraction of the original set.
\bpr
For each random line $L_j \in \m{L}^i$ all $N$ points from the $d$-dimensional space are projected onto the random line $L_j$, which takes time $O(dN)$. The number of random lines required until a point $P$ is in a set of size less than $minPts$ is bounded as follows: In each recursion the given set $\hat{S}$ is split into two sets $\hat{\m{S}}_0,\hat{\m{S}}_1$. By $p(E_{|\hat{\m{S}}|/4})$ we denote the probability of event $E_{|\hat{\m{S}}|/4}:= \min(|\hat{\m{S}}_0|,|\hat{\m{S}}_1|)\geq |\hat{\m{S}}|/4$ that the size of both sets is at least $1/4$ of the total set. As the splitting point is chosen uniformly at random, we have $p(E_{|\hat{\m{S}}|/4})=1/2$.
 Put differently, the probability that a point $P$ is in a set of size at most $3/4$ of the overall size $|\hat{\m{S}}|$ is at least $1/2$ for each random line $L$. 

 When projecting onto $|\m{L}^i|=c_1\cdot x$ lines for some value $x$, we expect $E_{|\hat{\m{S}}|/4}$ to occur  $c_1\cdot x/2$ times.  Using Theorem \ref{thm:Che} the probability that there are fewer than  $c_1\cdot x/4$ occurrences is $e^{-c_1\cdot x/48}$.
Using $x=\log N$ yields $e^{-c_1\cdot \log N/48}=1/N^{c_1/48}$ and for a suitable constant $c_1$ we have $N\cdot (3/4)^{c_1\cdot \log N/4} <1$. Therefore, the number of recursions until point $P$ is in a set $\hat{\m{S}}$ of size less than $minPts$ is at most $c_1\cdot \log N$ whp. Using Theorem \ref{thm:depEv} this holds for all $N$ points whp. A single projection takes time $O(dN)$. Thus, the time to compute $|\m{L}^i|=c_1\cdot \log N$  projections is $O(dN \log N)$ whp. 
\epr

Algorithm PerturbMultiPartitition calls Algorithm Partition $c_0\log N$ times; thus using Theorem \ref{thm:depEv}:

\medskip
\bco\label{co:multiPart}
Algorithm PerturbMultiPartitition runs in $O(dN(\log N)^2)$ time whp.
\eco

\section{Data Properties} \label{sec:prop}
\begin{figure}[htp]

\centerline{\includegraphics[width=0.4\linewidth]{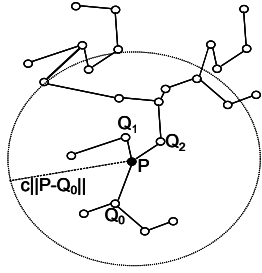}}

\caption{The parameter $B(P,c)$ used in the adaptive analysis is shown for a SLC and a point $P$ together with the points $N(P)=\{Q_0,Q_1,Q_2\}$ to which there is an edge in the SLC from $P$. $Q_0$ is the point furthest from $P$. $B(P,c)$ gives the points within a circle of radius $c\cdot ||P-Q_0||$ from $P$. The larger $|B(P,c)|$ the more points are projected onto random line $L$ between $P$ and $Q_i$. The minimum required size of a set $\m{S}$ grows with $|B(P,c)|$ to ensure that the endpoints of the edges $\{P,Q_i\}$ do not get partitioned into distinct sets. }
\label{fig:adaPara}
\end{figure}

We characterize the computational complexity, i.e. the ``difficulty'' of the data, by introducing a parameter $B$ to perform adaptive analysis. The parameter approximates the number of points that are expected to be projected between the two endpoints of an edge contained in the HC.  More precisely, the parameter $B(P,c)$ defines a set of points that for a point $P$ are within $c$ times the length of the longest edge  $e_l \in \m{HC}(P)$ (see Figure \ref{fig:adaPara}). The intuition is that all points in $B(P,c)$ are so close to $P$ that they have a ``rather large'' probability to end up in the same sets $\m{S}$ as $P$. 
\begin{definition}
$B(P,c):= \{Q \in \m{P}| D(P,Q) \leq c \cdot \max_{e'=\{C,D\} \in \m{HC}(P)} D(C,D)\}$ for some value $c$. 
\end{definition}

For a point $P$, the number of points within some distance $r$ increases monotonically with $r$. In general, the lengths of the edges vary significantly, i.e., if all edges are of similar length then it is not possible to find well-separated clusters. Thus the number of points $|B(P,c)|$ is expected to be small compared with the overall number of points $N$. 
The parameter $B(\m{P},c)$ for all points $\m{P}$ is given by the set $B(P,c)$ of maximum cardinality: $B(\m{P},c):=B(Q,c),$ s.t. $|B(Q,c)|=\max_{P \in \m{P}} |B(P,c)|$. 

\smallskip

\begin{figure}[!ht]

\centerline{\includegraphics[width=0.5\linewidth]{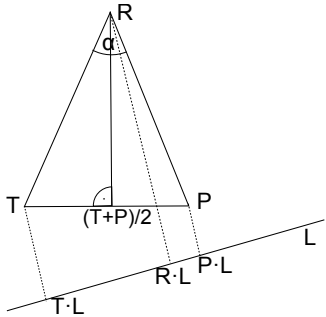}}

\caption{The probability that point ${R}$ is projected between ${T}$ and ${P}$ is given by the angle $\alpha$ divided by $\pi$.}
\label{fig:rightTria} 
\end{figure}

Next we compute the probability that for a random line a point $R$ is projected between $P$ and $T$ although it (potentially) is further away from both $P$ and $T$ than $T$ is from $P$ multiplied by a constant, i.e. $D(P,T)/ \leq  2\sin(1)\min(D({P},{R}),D({T},{R}))$, see Figure \ref{fig:rightTria}. We show that for the worst-case arrangement of three points $P,T,R$, the probability depends linearly on the ratio of the distances $D(P,T)$ and $D(P,R)$. 

\medskip
\begin{definition} \label{def:proj}
Let $PR(P,R,T)$ be the event that for a random line $L$ a point $R$ is projected between points $P$ and $T$, i.e., $(L\cdot {P} \leq  L \cdot {R} \leq L \cdot {T}) \vee (L\cdot {P} \geq  L \cdot {R} \geq L \cdot {T})$.
\end{definition}

\medskip
\bth\label{th:3pts} 
The probability $p(PR(P,R,T))  \leq \frac{D({P},{T})}{\pi D({P},{R})}$ given $D(P,T) \leq  2\sin(1)\min(D({P},{R}),D({T},{R}))$.
\ethe

\bpr 
The probability $p(PR(P,R,T))$ is given by $\alpha/\pi$, where the angle $\alpha$ is between vectors ${T}-{R}$ and ${P}-{R}$ (see Figure \ref{fig:rightTria}).  To maximize $\alpha$, we can assume that $R$ is as close to the point $(T+P)/2$ as possible. As by assumption $D({P},{T}) \leq  2\sin(1)D({P},{R})$ and $D(P,T) \leq  2\sin(1)D({T},{R})$, we assume that $D({P},{R})= D({T},{R}) =D(P,T)/(2\sin(1))$. In other words $T,R,P$ is a triangle with two sides of equal length.
We have $\sin(\alpha/2) =  \frac{D({P},{T})}{2\cdot D({P},{R})}\leq 2\sin(1)/2$. Thus, $\alpha/2\leq \sin^{-1}(2\sin(1)/2)\leq 1$. Therefore, looking at the series expansion of the sine function, we get $\sin(\alpha/2)=(\alpha/2)/1!-(\alpha/2)^3/3!+(\alpha/2)^5/5!+...\geq \alpha/2$, as $(\alpha/2)^i/i! \geq (\alpha/2)^{i+2}/(i+2)!$ because $\alpha/2\leq 1$. We have $\sin(\alpha/2) =  \frac{D({P},{T})}{2 D({P},{R})}\geq \alpha/2$. Thus $p(PR({P},{R},{T})) = \alpha/\pi \leq  \frac{D({P},{T})}{\pi D({P},{R})}$.
\epr

\section{Single Linkage Clustering and Minimum Spanning Tree}\label{sec:MST}
Using as input the outcome of the previous data-partitioning method,
we can compute the SLC or the MST using Algorithm \ref{alg:RPSLC}, which we call RP-SLC.
Data points are first partitioned using Algorithm PerturbMultiPartition without any perturbation, i.e., with parameter $l_{per.}=0$. After the partitioning, each point becomes a cluster. For each set of points $\m{S}$ resulting from the partitioning, we compute the distances of all pairs in the set. The union of all distances between any pair of points in any of the sets yields the set of all distances $\m{D}$. For SLC we iteratively merge the two clusters of smallest distance and update the remaining distances accordingly. Thus, the sequence of ascending distances $\m{D}_S$ forms the basis of merging two clusters. For the smallest (non-considered) distance computed $D(P,Q) \in \m{D}_S$, we check whether both points $P,Q$ belong to distinct clusters. If so, the clusters are merged. Clusters are merged repeatedly either until only one cluster remains or all pairs of points for which distances have been computed have been considered. As we do not take all pairwise distances among points $\m{P}$ into account but only distances among points from sets $\m{S} \in \mf{S}$, it is possible that we end up with more than one cluster. If that is the case, then too small a value has been chosen for parameter $minPts$. 

\begin{algorithm}[h!]
\caption{RP-SLC(points $\m{P}$, $minPts$) return $\m{HC}$ \label{alg:RPSLC}}
\begin{algorithmic}[1]
\begin{small}
\STATE $\m{HC}:=()$
\STATE  $\mf{S}:= $ Result of PerturbMultiPartition($\m{P},minPts,0$)
\FORALL{ $\m{S} \in \mf{S}$} 
	\FORALL{pairs $P, Q \in \m{S}$}	  
						\STATE $\m{D}:=\m{D} \cup D(P,Q)$ \label{alg:dist} \COMMENT{Compute distances}
	\ENDFOR
\ENDFOR
\STATE $\m{D}_S$:= Sorted distances $\m{D}$
\STATE $\m{C}:=\{Cl(P)| Cl(P)=\{P\},  P \in \m{P}\}$ \COMMENT{Initial clusters}
\WHILE{$|\m{C}|> 1 \wedge |\m{D}_S|>0 $}
	\STATE $\{P,Q\}:= \{R,T\} \in S,$ s.t. $D(R,T) = \min_{D(A,B) \in \m{D}_S} D(A,B)$  \COMMENT{Shortest edge}
	\STATE $\m{D}_S:=\m{D}_S \setminus D(P,Q)$
	\IF{$Cl_{ID}(P)\neq Cl_{ID}(Q)$}				
	 \STATE $\m{C}:=\m{C}\setminus \{Cl(P),Cl(Q)\} \cup (Cl(P) \cup Cl(Q))$ \COMMENT{Merge clusters $Cl(P)$ and $Cl(Q)$}
	 \STATE $\m{HC}:=\m{HC} \cup \{P,Q\}$ \COMMENT{Add edge to HC}
	 \ENDIF	
\ENDWHILE
\end{small}
\end{algorithmic}
\small
\smallskip
\end{algorithm}

\subsection{Analysis of the algorithms}

\bth \label{thm:SiComp}
Algorithm RP-SLC runs in $O(dN  \log N(minPts+  \log N))$ time whp.
\ethe

The runtime is dominated by the time to project all points and compute all relevant distances. 

\bpr
The PerturbMultiPartition algorithm takes time $O(dN \log^2 N)$ (see Corollary \ref{co:multiPart}) whp. Computing all pairwise distances for a set $\m{S} \in \mf{S}$ with all points $P \in \mathbb{R}^d$ takes time $O(d\cdot |\m{S}|^2)$. A set $\m{S}$ is of size $|\m{S}|< minPts$ and each point $P$ occurs in exactly $c_0 \log N$ sets $\m{S} \in \mf{S}$ since the Partition algorithm computes disjoint sets $\m{\hat{S}} \in \mf{\hat{S}}'$ and the PerturbMultiPartition algorithm returns the (mapped) union of them, i.e. $\mf{S}$. The time is maximized for $|\m{S}|= minPts-1$ resulting in $|\mf{S}|=c_0\log N \cdot N/(minPts-1)$. Therefore, the time for distance computations is $O( d \cdot N\cdot minPts\cdot \log N)$. A merger of clusters $Cl(P)$ and $Cl(Q)$ takes time proportional to the size of the smaller cluster, i.e., the time to update $Cl(P)$ as well as $Cl_{ID}(P)$ for points $P$ in the smaller cluster. There are at most $N-1$ mergers. The running time is maximal if both merged clusters are of the same size. Therefore, we have $N/2$ mergers of clusters of size 1, $N/4$ of clusters of size 2 and so on. Thus all merger operations together require at most $\sum_{i \in [0,\log N-1]} 2^i\cdot N/2^{i+1}= \sum_{i \in [0,\log N-1]}  N/2 = N/2 \log N-1$ time. The maximal number of iterations until all points have been merged into a single cluster (or all distances $\m{D}_S$ have been considered) is given by $O(N \cdot minPts\cdot \log N)$. Thus in total we get: $ O(dN \log^2 N)+ O(dN\cdot  minPts \cdot \log N) = O(dN  \log N(minPts+  \log N))$
\epr

Next, we prove in two steps that the computed SLC is correct, i.e., all relevant distances are taken into account. We require that both endpoints for every edge $e$ in the SLC be contained in one of the computed sets, i.e., the two points are not split. This, in turn depends on how many points are (expected to be) projected in between the two endpoints onto a random line and the maximal number of points in a set, i.e., $minPts$. For a single point, the likelihood to be projected in between depends on the distances to the endpoints of the edge. Thus, if there are not many points within a (short) distance of the endpoints relative to the maximum size $minPts$ of a set, the probability that both endpoints remain in the same set is large. Theorem \ref{thm:fracSets} quantifies this behavior by stating an upper bound on the number of projection sequences required such that each edge $e$ of length at most twice the longest edge $e_l \in SLC$ is contained in a set $\m{S} \in \mf{S}$ (see algorithm PerturbMultiPartition). This implies that all edges $e \in SLC$ are contained in a set $S$ and thus the HC is correct, see Theorem \ref{thm:ParaMST}. 

\medskip
\bth \label{thm:fracSets}
If $|B(\m{P},c_B)| \leq minPts/c_B$ with $c_B:=c_1\cdot \log^2 (N/minPts)\geq 1$ for a constant $c_1$ then whp for each edge $e=\{P,T\}$ with $D(P,T)\leq 2D(A,B)$ and $e_l=\{A,B\} \in SLC$ being the longest edge in the SLC, there exist at least $c_0\log N\cdot c_4(c_1)$ sets $\m{S}$ s.t. $e \in \m{S} \in \mf{S}$.
\ethe

The main idea of the proof is as follows: For two points $P,T$, we compute the number of points that are projected in between the projected points $P,T$ onto a random line $L$. We assume that all close points are projected in between them, but using Theorem \ref{th:3pts}, only a small fraction of far away points lie in between them. So, the number of points between them is limited to nearby points,  roughly points $B(\m{P},c_B)$. In conclusion, if the near points are less than (approximately) $B(\m{P},c_B)  \approx minPts$, the points $e=\{P,T\}$ remain in the same set $\m{S}$.

\bpr
Consider a random line $L$ onto which we project $|\m{S}|\geq minPts$ points, which are then split into two sets. Consider any edge $e=\{P,T\} \in \m{S}$ shorter than twice the length of $e_l$. Next, we use that for any point $R \in \m{P}\setminus B(P,c_B/2)$ we have by definition of $B(P,c_B/2)$ and $c_B$ that $\min(D(P,R),D(T,R))\geq D(P,T)$ and thus Theorem \ref{th:3pts} applies. Furthermore, we use the assumption  $|B(\m{P},c_B)| \leq  minPts/c_B$: 
 
\[ \begin{medsize}\begin{aligned}
 &E[PR(P,\m{P},T)]:=E[\sum_{R \in \m{P}} PR(P,R,T)] = 
 E[\sum_{\substack{ R \in B({P},c_B/2)\\ }} PR(P,R,T)]\\
       &  \text{\phantom{abc} }+ E[\sum_{\substack{R \in \m{P}\setminus\\ B({P},c_B/2)} } PR(P,R,T)|\min(D(P,R),D(T,R))\geq D(P,T)]\\
 &\leq  |B(\m{P},c_B/2)| + \frac{|\m{S}|}{c_B} \leq  2|\m{S}|/c_B  
 \end{aligned}\end{medsize} \]

Using Markov's inequality, the probability that the true value of $PR(P,\m{P},T)$ is larger than a factor $\sqrt{c_B}$ of the expectation  $E[PR(P,\m{P},T)]$ is bounded by $1/\sqrt{c_B}$. 

Consider Theorem \ref{thm:Part}. We require at most $O(\log N)$ projections until $|S|\leq minPts$ whp. This holds for all sequences $\m{L}^i \in \mf{L}$  whp using Theorem \ref{thm:depEv}. Thus the probability that after at most $c_2\log N$ projections $PR(P,\m{P},T)$ is larger by a factor of at most $\sqrt{c_B}$ of its expectation is bounded by \[ \begin{medsize}\begin{aligned} (1-1/c_B)^{c_2\log N}= (1-1/(\sqrt{c_1}\log N))^{c_2\log N} = 1/e^{c_2/\sqrt{c_1}}\end{aligned}\end{medsize}\]
Therefore, with probability $1/e^{c_2/\sqrt{c_1}}$, the following event $E_0$ occurs for all sequences $\m{L}^i \in \mf{L}$: $PR(P,\m{P},T)\leq 2|S|/\sqrt{c_B}$. Assume $E_0$ occurred. Then for each edge $e=\{P,T\}$ there are at most $2|S|/\sqrt{c_B}$ projected points between $P$ and $T$ for a set $\m{S}$. The probability $p(SP(P,T))$ that points $P,T$ end up in different sets after a set $\m{S}$ is being projected onto line $L$, i.e. $P \in \m{S}_0$ and $Q \in \m{S}_1$ with $\m{S}_0 \cup \m{S}_1 = \m{S}$ is maximized if the number of points being projected between $P$ and $Q$ on line $L$ is maximized. If for a single projection all $2|S|/\sqrt{c_B}$ points as well as $P$ and $T$ belong to the same set  then $p(SP(P,T)) \leq 2|S|/\sqrt{c_B}/|\m{S}| = 2/\sqrt{c_B}$. Define event $E_1|E_0$ that for all projections $L \in \mf{L}_i, \exists \m{W} \in \mf{W}, \text{ s.t. } P,T \in \m{W}$ given that $E_0$ occurred. It holds for a suitable constant $c_1$ that
\[ \begin{medsize}\begin{aligned} &p(E_1|E_0) \geq (1-p(SP(P,T)))^{c_2\log N} \\& = (1-2/(\sqrt{c_1}\log N))^{c_2\log N} \geq 1/e^{2c_2/\sqrt{c_1}} \end{aligned}\end{medsize} \]
The probability that $E_0$ and $E_1|E_0$ occur can be bounded by \[p(E_1|E_0)\cdot p(E_0) \geq  1/e^{4c_2/\sqrt{c_1}}=c_5(c_1)\] 

Thus, for $|\mf{L}|=c_0 \log N$ sequences of random projections $\m{L}^i \in \mf{L}$ for an arbitrary constant $c_0$ using Theorem \ref{thm:Che} with probability $1-e^{c_0\log N\cdot c_5(c_1)}$ for a constant $c_5(c_1)$, there exist $\log N c_4(c_1)$ sets $\m{S} \in \mf{S}$ containing both end points $P$ and $T$ of an edge $e=\{P,T\} \in SLC$, i.e. $P,T \in \m{S}$. Using Theorem \ref{thm:depEv}, this holds for all $N-1$ edges $e \in SLC$.
\epr

\medskip
\bth \label{thm:ParaMST}
When $|B(\m{P},c_1 \log^2 (N/minPts))| \leq minPts/(c_1\log^2 (N/minPts))$, an SLC is computed in time $O(N\cdot minPts\log N(d+  \log N))$ whp.
\ethe
Note: The term whp refers to both runtime and correctness.
\bpr
Because of Theorem \ref{thm:fracSets} each edge $e\in SLC$ occurs in multiple sets $\m{S} \in \mf{S}$ whp. Thus it is considered by the computation. The complexity is bounded by Theorem \ref{thm:SiComp}.
\epr

\subsection{Parameter-free RP-SLC algorithm} \label{sec:NoPara} 
The RP-SLC algorithm described in the preceding section requires as input a parameter $minPts$. The parameter states the maximal size of a set of points so that it is not further partitioned. We face the following trade-off: Choosing the parameter $minPts$ too large has a negative impact on the running time; setting it too small may result in an incorrect clustering because of edges that are ignored. The goal is therefore to find the smallest value for $minPts$ such that the clustering is correct. There are two conditions that make it likely that an edge in the SLC is not considered:

(i) The shortest edge $e$ between two clusters found in any of the sets (computed by algorithm PerturbMuliPartition) occurs only in a few sets. In this case, there is a good chance that there exists an even shorter edge $e'$ that is not present in any set at all. If this edge $e'$ is in the SLC, it will be missed.

(ii) There are many (nearly) colinear points with line $P$ to $Q$ as shown in Figure \ref{fig:densityperturb}. These points are likely to be projected between the end points of an edge $e_i=\{P,T\}$ in the SLC, but unlikely to be projected onto the edge $e=\{P,Q\}$ not in the SLC. Therefore, the algorithm picks the wrong edge $\{P,Q\}$ for merging clusters because $P$ and $T$ end up in distinct sets.  

To deal with conditions (i) and (ii), we extend the RP-SLC algorithm. Our parameter-free RP-SLC algorithm finds the smallest parameter $minPts$ asymptotically. Roughly speaking, we partition points and merge clusters (as for the RP-SLC algorithm), but repeat this process for different parameters $minPts$ to ensure that no edge in the SLC is missed by the algorithm. 

\medskip
\noindent \underline{Condition (i):} To deal with Condition (i),  we make sure that the currently shortest edge $e$ considered by the algorithm is frequent, i.e., it occurs in a certain fraction of all sets.  This guarantees that it becomes extremely unlikely that an even shorter edge $e'$ exists that does not occur in any set at all. A frequent edge can be either between two different clusters or an intercluster edge $e'$. (An intercluster edge $e'$ is an edge between two points of the same cluster.) To maintain the merging process, we require that there be at least one feasible edge for each point. An edge is feasible if it is frequent and connects two distinct clusters. If a point $P$ is not adjacent to a feasible edge, then the longest frequent intercluster edge $e'$ adjacent to $P$ has to be longer than the shortest feasible edge of all points. This ensures that although there might be shorter edges than $e'$ adjacent to $P$, these shorter edges are almost certainly not shorter than the currently shortest edge considered by the algorithm.

 
\begin{figure}[!ht]
\centerline{\includegraphics[width=0.5\linewidth]{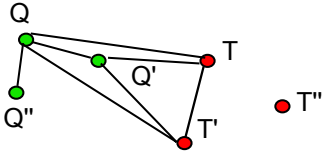}}

\caption{Consider clusters $\{Q,Q',Q''\}$ and $\{T,T',T''\}$ and sets $S_0=\{Q,Q',T\}$, $S_1=\{Q,Q',T,T'\}$ and $S_2=\{Q'',Q\}$. Requiring that an edge is frequent if it occurs in 2/3 of all sets, i.e. $c_f=2/3$, we have that edges $\{Q',Q\}$, $\{Q',T\}$ and $\{Q,T\}$ are frequent. $\{Q',Q\}$ is an intercluster edge, i.e., taken. $\{Q,T\}$ is feasible, i.e., usable for merging.} 
\label{fig:defs} 
\end{figure}

More formally, an edge $\{Q,P\}$ is \emph{frequent} for $P$ if $\{Q,P\} \in \m{S}$ for a fraction $c_f$ of all sets $\m{S}$ containing $P$. An example is shown in Figure \ref{fig:defs}.  To determine whether an edge is frequent, we compute the number of times $n(P,Q)$ a pair $P,Q$ of points, i.e. an edge $\{P,Q\}$, has occurred in any set $S \in \m{S}$.
For each point $P$, we compute the feasible edges to points $Q \in \m{F}(P)$ and the taken edges to points $Q \in \m{T}(P)$ defined next.  An edge $\{Q,P\}$ is \emph{feasible} if it is frequent and $Q$ and $P$ belong to different clusters. An edge $\{Q,P\}$ is \emph{taken} if it is frequent and $Q$ and $P$ belong to the same cluster. Thus, the feasible and the taken edges change after every merger. We require that there be at least one feasible edge, i.e. $|\m{F}|>0$ (see Condition \ref{eq:cond}).  Every point $P$ in some set $\m{P}^{check}$ (defined below) has to be adjacent to a feasible or a taken edge. If it is only adjacent to a taken edge, then the maximal length of a taken edge must be larger than the shortest feasible edge of any point: 
\[  \begin{medsize}
\begin{aligned}  
|\m{F}|>0 \wedge & \forall P \in \m{P}^{check}:  |\m{F}(P)|>0 \vee  \label{eq:cond} \phantom{sdfasfasdabcdefa} \text{(6.2)} \phantom{abcdefafdfm}  \\ & \left(|\m{T}(P)|>0 \wedge \max_{T \in \m{T}(P)} D(P,T)\geq 
 \min_{Q \in \m{P},R \in \m{F}(Q)} D(Q,R)\right) 
\end{aligned}
 \end{medsize}\]
Before the first merger of two clusters, we check Condition \ref{eq:cond} for all points, $\m{P}^{check}= \m{P}$. After the $i$-th merger due to edge $e_i=\{P,Q\}$, we only consider points that are contained in a set $\m{S}$ containing both $P$ and $Q$. If Condition \ref{eq:cond} is violated, we double $minPts$ until (\ref{eq:cond}) is fulfilled. \smallskip \\ 

\noindent
\underline{Condition (ii):} To deal with Condition (ii),  points are perturbed, see Figure \ref{fig:densityperturb}.
Controlled perturbations are used in computational geometry \cite{Meh11}. We share the idea that the input is modified in a well-defined manner. Whereas for controlled perturbations an exact output is computed for a perturbed input, our final outcome is an exact solution for the unperturbed input. Our perturbation scheme ensures a more equal distribution of points and thus avoids the scenario that the number of points projected onto edge $e$ and $e'$ adjacent to the same point differs by more than a constant factor.\footnote{It might be possible to do without perturbations using a larger parameter $minPts$ and explicitly detecting cases like Figure \ref{fig:densityperturb}.} 
 More precisely, onto any edge $\{P,Q\}$ adjacent to a point $P$ that is larger than the shortest edge $\{P,T\}$ adjacent to point $P$ roughly the same number of points (or more) are projected. Therefore, if any edge $\{P,Q\}$ occurs frequently, i.e., its end points are contained in many sets, then also $\{P,T\}$ must occur frequently as well. Thus, it suffices to raise $minPts$ until the shortest adjacent edge $\{P,Q\}$ adjacent to $P$ that is considered by the algorithm is frequent. We can do this in the same manner as for Condition (i), i.e., by ensuring that Condition \ref{eq:cond} is satisfied.

 It remains to discuss how strongly the points should be perturbed. To compute a SLC, we iteratively consider (feasible) edges of increasing lengths. Perturbing the points with a vector of length $l_{per.}$ proportional to the length of the currently investigated feasible edge is sufficient to deal with artificial distributions of points as in Figure \ref{fig:densityperturb}. Therefore, we have to adapt the length $l_{per.}$ of the perturbation vector and perturb all points again and again.
More precisely, once the currently shortest feasible edge $e'$ is 1/8 of $l_{per.}$ we set $l_{per.}$ to be 1/16 of the length of $e'$ and recompute the partitioning of points (using the PerturbMultiPartition algorithm). As we shall see, in this case (nearly) colinear points close to a point $P$ are perturbed sufficiently such that they are somewhat likely to be projected onto any edge $e$ with $P \in e$ and not just onto the shortest edge adjacent to $P$. 

\begin{algorithm}[h!]
\caption{Parameter-Free RP-SLC(points $\m{P}$) return $\m{HC}$ \label{alg:ParaFree}}
\begin{algorithmic}[1]
\begin{small}
\STATE $\m{HC}=()$, {$l_{per.}:=0$} \label{par:a}, {$minPts:=c_0 \log N$}   
\STATE $\forall P,Q \in \m{P}: n(P,Q):=0$
\REPEAT
\STATE $\mf{S}:= $ { Result of PerturbMultiPartition}($\m{P},minPts,l_{per.}$) \label{par:b}
\FORALL{ $\m{S} \in \mf{S}$} 
	\FORALL{pairs $P, Q \in \m{S}$}	  						
						\STATE $\m{D}:=\m{D} \cup D(P,Q)$ \COMMENT{Compute distances}
						\STATE $n(P,Q):=n(P,Q)+1$ \COMMENT{Count occurences of edges}
\ENDFOR
\ENDFOR
\STATE $\m{F}(P):=(Q \in \m{P}| Cl_{ID}(P)\neq Cl_{ID}(Q) \wedge \frac{n(P,Q)}{c_0\log N} > c_f)$ sorted according to distance to $P$ for constant $c_1$
\STATE $\m{T}(P):=(Q \in \m{P}| Cl_{ID}(P)= Cl_{ID}(Q) \wedge \frac{n(P,Q)}{c_0\log N} > c_f)$ sorted according to distance to $P$
\STATE $\m{P}^{check}:=\m{P}$
\STATE $\m{C}:=\{Cl(P)| Cl(P)=\{P\},  P \in \m{P}\}$ \COMMENT{Initial clusters}
\WHILE{$|\m{C}|> 1 \wedge $ \text{ Condition \ref{eq:cond}} $\wedge \min_{D(A,B) \in \m{F}} D(A,B)/8 \geq l_{per.}$}  \label{par:c}
	\STATE $(P,Q):=$ shortest edge $e \in \m{D}$  
	\IF{$Cl_{ID}(P)\neq Cl_{ID}(Q)$}				
	 	 \STATE $\m{C}:=\m{C}\setminus \{Cl(P),Cl(Q)\} \cup (Cl(P) \cup Cl(Q))$ \COMMENT{Merge clusters $Cl(P)$ and $Cl(Q)$}
		 \STATE $\m{P}^{check}:=\m{F}(P) \cup \m{F}(Q)$	 	
		 \STATE $\m{HC}:=\m{HC} \cup \{P,Q\}$
	 \ENDIF	
	 \ENDWHILE
	 \STATE \textbf{if } not Condition \ref{eq:cond} \textbf{ then }  $minPts:=2minPts$ \textbf{ end}
	 \STATE {$l_{per.}:= \min_{D(A,B) \in \m{F}} D(A,B)/16$} \label{par:d}
\UNTIL{$|\m{C}|=1$}
\end{small}
\end{algorithmic}
\small
\smallskip
\end{algorithm}





To show that the algorithm works correctly, we must prove that before the $i$-th merger the edge $e_i=\{T,P\} \in \m{HC}$ of the SLC also occurs in one of the sets $\mf{S}$ considered, i.e. $\exists \m{S} \in \mf{S}$, s.t. $\{T,P\} \in \m{S}$. We prove that because of the perturbation of points this holds if Condition \ref{eq:cond} is satisfied. As our algorithm increments $minPts$ until Condition \ref{eq:cond} is satisfied, to prove this it suffices to show correctness of the algorithm.  

\smallskip

\ble
Before every merger there is a feasible edge, i.e., $|\m{F}|>0$, and for every point $P \in \m{P}$ there exists at least one feasible or at least one taken edge, i.e., $|\m{F}(P) \cup \m{T}(P)|>0$. 
\ele

\bpr
If there are no feasible edges ($|\m{F}|=0$) or  a point $P \in \m{P}$ has neither a feasible nor a taken edge, i.e. $|\m{F}(P)|=|\m{T}(P)|=0$, then owing to the while Condition (line \ref{par:c} in Algorithm \ref{alg:ParaFree}), the algorithm does not merge any cluster but doubles $minPts$. For $minPts\geq N$ all pairwise distances among all points are considered.
\epr

Next, we define an event $X_{\hat{P},\hat{Q},\hat{T}}$ that ensures that the projection of the (perturbed) shortest edge $\{T,P\} \in SLC$ onto an edge $\{P,Q\}$ is of some minimum length. In turn, this guarantees that any point projected between $\hat{P},\hat{T}$ is also likely to be projected onto edge $\hat{P},\hat{Q}$ (see Figure \ref{fig:densityperturb}). Lemmas \ref{thm:pro} and \ref{le:singlePro} quantify this. This in turn allows us to compute the probability that the endpoints $T,P$ of the edge $\{T,P\}$ are split for a projection relative to the probability that the endpoints $P,Q$ of the edge $\{P,Q\}$ are split (Lemma \ref{le:worstcase2}).

Let event $X_{\hat{P},\hat{Q},\hat{T}}$ be the event that  $|(\hat{P}-\hat{Q})\cdot (\hat{T}-\hat{P})|\geq D(T,P)/32$ given $D(P,Q)\geq D(P,T)$.

\ble \label{thm:pro}
The probability of event $X_{\hat{P},\hat{Q},\hat{T}}$ is 2/3.
\ele
The proof uses a worst-case arrangement of points $P,Q,T$ before perturbation. 

\bpr
We use a worst-case arrangement of points $P,Q,T$ before perturbation to compute a bound on the probability. To minimize the dot product $|(\hat{P}-\hat{Q})\cdot (\hat{T}-\hat{P})|$, we assume that the unperturbed points form a right triangle, i.e. the dot product $({P}-{Q})\cdot ({T}-{P})=0$. Any non-zero angle only increases the dot product of the perturbed points, since it introduces a bias. With the same reasoning we assume that even after the perturbation of $P$ and $T$, the points $P$ and $T$ form a rectangular triangle, i.e. the dot product $(\hat{P}-{Q})\cdot (\hat{T}-\hat{P})=0$. Due to our assumptions, only the perturbation vector $\hat{Q}-Q$  contributes to increase the dot product. We get for all 3 perturbed points: $(\hat{P}-\hat{Q})\cdot (\hat{T}-\hat{P})=(\hat{P}-\hat{Q}+Q-Q)\cdot (\hat{T}-\hat{P})=(\hat{P}-Q)\cdot (\hat{T}-\hat{P})+(-\hat{Q}+Q)\cdot (\hat{T}-\hat{P})=(-\hat{Q}+Q)\cdot (\hat{T}-\hat{P})=\cos(\theta)\cdot |\hat{T}-\hat{P}|\cdot |Q-\hat{Q}|$. To minimize $|\hat{T}-\hat{P}|\cdot |Q-\hat{Q}|$ we can maximize the angle $\theta \in[0,\pi/2]$, i.e. minimize $|\hat{T}-\hat{P}|$ and $|Q-\hat{Q}|$. Since each point is perturbed by a vector of length $D(T,P)/8$ we assume that the vector ${T}-{P}$ is shortened owing to perturbation to $|\hat{T}-\hat{P}| = 7D(T,P)/8$. For $|Q-\hat{Q}|$ we use the lower bound due to the algorithm, i.e. $D(T,P)/16$. Since the perturbation vector is chosen uniformly at random, any angle $\theta \in [0,\pi]$ has the same probability. Thus, in particular the probability $\theta \in ([0,\pi/3] \cup  [2\pi/3,\pi])$ is 2/3. For any of the values $([0,\pi/3] \cup  [2\pi/3,\pi])$ we have $|\cos(\theta)|\geq 0.5$. Therefore the length is $D(T,P)/32$ with probability 2/3.  
\epr

\ble \label{le:singlePro} 
Given $X_{\hat{P},\hat{Q},\hat{T}}$ , we have $p(PR(\hat{P},\hat{R},\hat{Q}))/p(PR(\hat{P},\hat{R},\hat{T})) \geq (\arctan(161)-\arctan(160))$, independently of how other points $\hat{P}\setminus\hat{R}$ are projected.
\ele


\begin{figure}[!ht]
\centerline{\includegraphics[width=0.95\linewidth]{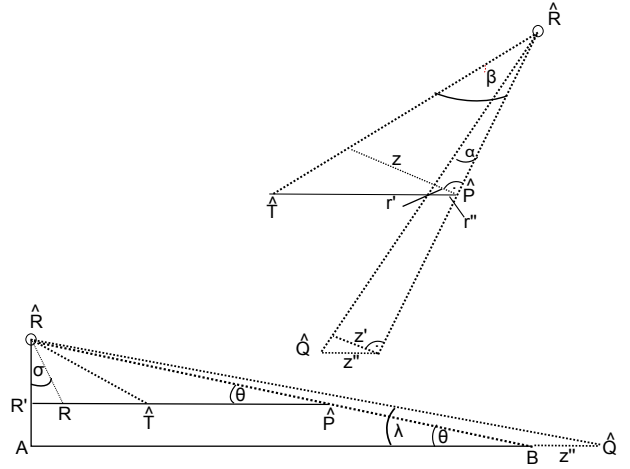}}

\caption{Illustration of the terms in the proof of Lemma \ref{le:singlePro}. The upper panel illustrates the case $\min(d(R,P),d(R,T))\geq 2D(T,P)$ and the lower panel $\min(d(R,P),d(R,T))< 2D(T,P)$.}
\label{fig:triaAngle} 
\end{figure}


The intuition of the proof is as follows: If $\hat{R}$ is close to the line through $\hat{P}$ and $\hat{T}$, then there is a constant probability for $PR(\hat{P},\hat{R},\hat{Q})$ because of event $X_{\hat{P},\hat{Q},\hat{T}}$. The same reasoning applies if $\hat{R}$ is far from the line through $\hat{P}$ and $\hat{T}$, i.e.  the triangles  $\hat{P},\hat{R},\hat{Q}$ and  $\hat{P},\hat{R},\hat{T}$  are constrained due to event $X_{\hat{P},\hat{Q},\hat{T}}$ such that points are projected not only between $\hat{P}$ and $\hat{T}$ but also between $\hat{P}$ and $\hat{Q}$. Figure \ref{fig:densityperturb} illustrates this. 

\bpr
Assume $\min(d(R,P),d(R,T))\geq 2D(T,P)$. 
The probability $p(PR(\hat{P},\hat{R},\hat{T}))$ is given by $\beta/\pi$ (see upper panel in Figure \ref{fig:triaAngle})  and in the same manner $p(PR(\hat{P},\hat{R},\hat{Q}))=\alpha/\pi$.
To bound the ratio $\alpha/\beta$ we first bound the ratio $z'/z$. Using the triangle inequality, we  have $D(\hat{R},\hat{Q})\leq D(\hat{Q},\hat{P})+D(\hat{P},\hat{R})$. Since $D(T,P)\leq D(P,Q)$ and the distortion vector of a single point is of length at most $D(T,P)/8$ we get $D(\hat{Q},\hat{P})\leq (1+1/4) D(T,P)$. With the same reasoning $D(\hat{P},\hat{R})\geq D(P,R)-1/4 D(T,P)\geq (2-1/4)D(T,P)$. Thus $D(\hat{R},\hat{Q})\leq 2\cdot D(\hat{R},\hat{P})$. By assumption  $X_{\hat{P},\hat{Q},\hat{T}}$ occured, therefore  we have  $z'':=|(\hat{P}-\hat{Q})\cdot (\hat{T}-\hat{P})|\geq D(T,P)/32$ (see Figure \ref{fig:triaAngle}), i.e. we use $z'':= D(T,P)/32$. We have $r'/r''= z/D(\hat{Q},\hat{P})$. Due to  $\min(d(R,P),d(R,T))\geq 2D(T,P)$ and $D(P,Q)\leq D(P,T)$ we have $D(\hat{P},\hat{Q})\leq D(\hat{P},\hat{R})$. Therefore $z''\leq 2r''$. Furthermore, $z'\geq r' $. Thus $z'/z'' \geq r'/(2r'')$. Therefore, $2z'/z'' \geq r'/r'' = z/D(\hat{Q},\hat{P})$. Using $D(\hat{Q},\hat{P}) \leq (1+1/4)D(T,P)$ and $z''\geq D(T,P)/32$, we get $2z'/(D(T,P)/32) \geq z/(1+1/4)D(T,P)$ and $z'\geq 1/90 z$.
 Also: $\tan(\beta)\geq \frac{z}{D(\hat{R},\hat{P})}$ and since $D(\hat{P},\hat{Q})\leq D(\hat{P},\hat{R})$ we get $\tan(\alpha)\geq \frac{z'}{2\cdot D(\hat{R},\hat{P})} = \frac{z}{180 D(\hat{R},\hat{P})}$.  Thus, setting for readability $y:=\frac{z}{ D(\hat{R},\hat{Q})}$, we get $\alpha\geq \frac{z}{180 D(\hat{R},\hat{Q})}$. Therefore $\alpha/\beta = \arctan(y/180)/\arctan(y)$. We have $\lim_{y \rightarrow \infty} \arctan(y/180)/\arctan(y)= 1$ and  $\lim_{y \rightarrow 0} \arctan(y/180)/\arctan(y)= 1/180$. The latter follows from a series expansion at $y=0$. For $y \in [0,1]$ because of the strictly monotonic increase of $\arctan(y)$, we also get a strictly monotonic increase of $\arctan(y/180)/\arctan(y)$. Therefore $\alpha/\beta\geq 1/180$.

Assume $\min(d(R,P),d(R,T))< 2D(T,P)$  (see lower panel in Figure \ref{fig:triaAngle}).  This implies $\max(d(R,P),d(R,T)) < 3D(T,P)$. We use the trivial bound $p(PR(\hat{P},\hat{R},\hat{T}))\leq 1$, i.e. to minimize the ratio $p(PR(\hat{P},\hat{R},\hat{Q}))/p(PR(\hat{P},\hat{R},\hat{T}))$ we use $p(PR(\hat{P},\hat{R},\hat{T}))= 1$. To bound $p(PR(\hat{P},\hat{R},\hat{Q}))$, we assume that $\hat{P}, {R},\hat{Q}$ are on the same line and that $D(\hat{Q},\hat{R})$ is maximal: $D(\hat{Q},\hat{R})\leq  D(\hat{Q},\hat{P})+ D(\hat{P},\hat{R})\leq (1+1/4)D(T,P)+(3+1/4)D(T,P) \leq 5 D(T,P)$. Next, we bound the distance $\hat{R}$ to the line $\hat{P}$ through $\hat{T}$, i.e. $D(\hat{R},R')$. We have $|(R-\hat{R})\cdot (\hat{R'}-\hat{R})|\geq  \cos(\sigma) |(R-\hat{R})|| (\hat{R'}-\hat{R})|$. With probability $2/3$ $\sigma \in [0,\pi/2]$ is inbetween $[0,\pi/3]$. Therefore, $|\cos(\sigma)|\geq 1/2$. Since $|(R-\hat{R})|\geq D(T,P)/16$ we get $|(R-\hat{R})\cdot (\hat{R'}-\hat{R})|\geq D(T,P)/32$.

To minimize $\lambda-\theta$, we can minimize $\theta$ (Figure \ref{fig:triaAngle}). For the triangle $A,B,\hat{R}$, we get that $d(A,\hat{R})\geq d(R',\hat{R})\geq D(T,P)/32 $ and $d(A,B)\leq 5D(T,P)$ $\theta \geq \arctan(d(A,\hat{R})/d(A,B))\geq (\arctan(5D(T,P)/ (D(T,P)/32))=\arctan(160)$. As $z'':=|(\hat{P}-\hat{Q})\cdot (\hat{T}-\hat{P})|\geq D(T,P)/32$ we get $\lambda \geq \arctan((5D(T,P)+z'')/ (D(T,P)/32))=\arctan((1+1/160)(5D(T,P))/ (D(T,P)/32))=\arctan(161)$. Thus $\lambda-\theta= \arctan(161)-\arctan(160)$.  Therefore, $p(PR(\hat{P},\hat{R},\hat{T}))/p(PR(\hat{P},\hat{R},\hat{Q})) \geq \arctan(161)-\arctan(160)$. 
\epr


Let $SP(A,B)$ be the event that for a single projection $L$ of a set $\m{S}$ with $A,B \in \m{S}$ the two points are split into different sets, eg. $A \in \m{S}_0$ and $B \in \m{S}_1$ with $\m{S}_0 \cup \m{S}_1 = \m{S}$.

\ble \label{le:worstcase2}
$\frac{p(SP({P},{Q}))}{p(SP({P},{T}))} \geq (\arctan(161)-\arctan(160))$  given $X_{\hat{P},\hat{Q},\hat{T}}$. 
\ele 
The proof relies on Lemma \ref{le:singlePro} to compute the number of expected points  projected between $P,Q$ and $P,T$, respectively.

\bpr
Let $Y(P,Q)$ denote the random variable stating the number of projected points from the set $\m{S}$ between $P,Q$ on a randomly chosen line $L$.
If there are no points projected between $P,Q$ and $P\cdot L < Q\cdot L$ and $P$ is a splitting point (or vice versa) then the $P$ and $Q$ end up in different sets. If any other point from $\m{S}$ is chosen they stay in the same set. Mathematically speaking, we have $p(SP(P,Q)|Y(P,Q)=0)=1/|\m{S}|$. More generally, if there are $x$ points between $P,Q$ we get $p(SP(P,Q)|Y(P,Q)=x)=(x+1)/|\m{S}|$.
\[ \begin{medsize}\begin{aligned}
p(SP(P,Q)) =\sum_{i=0}^{|\m{S}|-2} p(Y(P,Q)=i)\cdot p(SP(P,Q)|Y(P,Q)=i)\\
  =\sum_{i=0}^{|\m{S}|-2} p(Y(P,Q)=i)\cdot (i+1)/|\m{S}|\\
  =1/|\m{S}|\cdot (1+\sum_{i=0}^{|\m{S}|-2} p(Y(P,Q)=i)\cdot i)\\
\end{aligned} \end{medsize} \]
By definition the number of expected points $E[Y(P,Q)]=\sum_{i=0}^{|\m{S}|-2} p(Y(P,Q)=i)\cdot i$ projected between $P,T$ is also $E[Y(P,Q)]=\sum_{R \in \m{S}\setminus \{P,T\}} p(PR(P,R,T))$. Since a splitting point is chosen randomly among the $\m{S}$ points (See Algorithm Partition) we get $p(SP({P},{T}))= 1/|\m{S}|\cdot (1+\sum_{R \in \m{S}\setminus \{P,T\}} p(PR(P,R,T)))$.
Additionally, $p(PR(\hat{P},\hat{R},\hat{Q})) / p(PR(\hat{P},\hat{R},\hat{T})) \geq (\arctan(161)-\arctan(160))$  using Theorem \ref{le:singlePro}.

\[ \begin{medsize}\begin{aligned}
p(SP({P},{Q}))= 1/|\m{S}|\cdot (1+\sum_{R \in \m{S}\setminus \{P,Q\}} p(PR(P,R,Q))) \geq \\ 
1/|\m{S}|\cdot (1+ (\arctan(161)-\arctan(160))\sum_{R \in \m{S}\setminus \{P,T\}} p(PR(P,R,T))) \\
\\ \geq (\arctan(161)-\arctan(160))/|\m{S}|\cdot (1+ \sum_{R \in \m{S}\setminus \{P,T\}} p(PR(P,R,T))) \\
= (\arctan(161)-\arctan(160))\cdot p(SP({P},{T}))\end{aligned} \end{medsize} \]

Therefore, $\frac{p(SP({P},{Q}))}{p(SP({P},{T}))} \geq (\arctan(161)-\arctan(160))$ 
\epr

\bth \label{thm:RPSLCfree}
The parameter-free RP-SLC algorithm has time complexity $O(dN\log N \cdot (|B(\m{P},c_B)| +  \log N))$ for $c_B:=c_1\log^2 (N/minPts^*)$ and space complexity $O(dN+ N\log N \cdot (|B(\m{P},c_B)| +  \log N))$ whp.
\ethe

The proof uses Lemma \ref{le:worstcase2} to show that if an edge $\{P,T\}$ is not considered then any other edge $\{P,Q\}$ cannot be feasible and therefore $minPts$ is increased. Furthermore, using Theorem \ref{thm:fracSets} we obtain the bound for $c_B$.

\bpr
The space complexity can be upper bounded by the number of distances that are computed and the space to store all $N$ points.
Using Lemma \ref{thm:pro}, we have that $X_{\hat{P},\hat{Q},\hat{T}}$  is $2/3$. Using Theorem \ref{thm:Che}, $X_{\hat{P},\hat{Q},\hat{T}}$ occurs for at least a fraction of $c_0\log n\cdot 2/3$ whp.
Assume $X_{\hat{P},\hat{Q},\hat{T}}$ occurred. Consider a sequence of projections $\m{L}^i$ with $|\m{L}^i|\geq c_5$ for a constant $c_5$. Let us compute the probability that an edge $e=\{P,T\}\in SLC$ is not considered, i.e., there is no set $S \in \m{S}$ containing both points $P,T$ resulting from $\m{L}^i$.  Using Lemma \ref{le:worstcase2} we have for a single projection $L \in \m{L}^i$: $\frac{p(SP({P},{Q}))}{p(SP({P},{T}))} \geq c_6$ for a constant $c_6$. 

Thus for all $|\m{L}^i|$ projections we get as a bound for $p(SP(\hat{Q},\hat{P}))$: $(1-p({SP({P},{Q})}))^{|\m{L}^i|} \geq (1-c_6 p({SP({P},{T})}))^{|\m{L}^i|} \geq   (1-c_6)^{|\m{L}^i|}\geq (1-c_6)^{c_5} = 1/c_7$. Using Theorem \ref{thm:Che}  there are at least $c_0\log N\cdot 2/(3c_9) = c_0\log N/c_{10}$ sequences for (suitable) constants $c_0,c_9,c_{10}$ whp such that ${SP({Q},{T})}$ occurs. By setting $c_f=1-c_0/c_{10}$ in the algorithm the edge $Q,P$ is neither feasible nor taken whp. Thus, $minPts$ will be increased.

Using Theorem \ref{thm:fracSets} for $|B(\m{P},c_B)| \leq minPts^*/c_B$, there exists at least $c_0\log N\cdot c_2(c_1)$ projection sequences $\m{L}^i$ s.t. $e \in S \in \m{S}$  for each edge $e\in SLC$ whp. As each edge $e \in SLC$ is feasible if there exist $c_0\log N c_f$ such projection sequences, i.e., $n(e)\geq c_f$, all edges $e \in SLC$ will be feasible when $c_f\leq c_2(c_1)$.

Let $e'$ be the shortest edge adjacent to $P$ that is not in the SLC, i.e. $P \in e'$ and $e' \notin SLC$. The length of edge $e'$ is bounded by twice the size of the longest edge in the SLC. Using Theorem \ref{thm:fracSets} there exist at least $c_0\log N c_4(c_1)$ projection sequences such that edge $e'$ is either feasible or taken whp.

Owing to Theorem \ref{thm:SiComp} for a fixed number of $minPts$ the algorithm takes time $O(dN  \log N(minPts+  \log N))$.
Owing to Theorem \ref{thm:ParaMST} once $|B(\m{P},c_B)| \leq minPts^*/c_B$ the algorithm terminates computing a correct SLC  whp. The total runtime is whp $\sum_{i=1}^{\log (minPts^*/c_B)} O(N\cdot 2^i\log N(d+  \log N)) = O(dN\log N\cdot( minPts^*/c_B+  \log N)) = O(dN\log N \cdot (|B(\m{P},c_B)| +  \log N))$. 
\epr

\section{Average Linkage Clustering} \label{sec:Avg} 
For ALC, the two clusters $C_0,C_1 \in \m{C}$ are merged with minimum average (squared) distance of all pairs of points from distinct clusters, i.e. $D_{ALC}(C_0,C_1):=\sum_{P \in C_0,Q \in C_1} D(P,Q)^2/(|C_0||C_1|)$. However, it is known \cite{Lei06} (or Chapter 8 of \cite{Vip05}) that it is not necessary to explicitly compute the distances among all pairs, but it suffices to maintain the cluster centers and their variances. For a cluster $C_0$ the center is given by  $\mu(C_{0}) := \sum_{P \in C_0} P/|C_0|$ and the variance by $\sigma^2(C_{0}) := \sum_{P \in C_0} D(P,\mu(C_{0}))^2/|C_0|$. The \emph{ALC-distance} between two clusters $C_0,C_1$ becomes 

\begin{align} 
D_{ALC}(C_0,C_1)&:=\sum_{P \in C_0,Q \in C_1} \frac{D(P,Q)^2}{|C_0||C_1|}  \label{eq:alc} \\&= D(\mu(C_{0}),\mu(C_{1}))^2 + \sigma^2(C_{0})+ \sigma^2(C_{1}) \nonumber 
\end{align}

We can compute the new cluster center and variance incrementally after a merger:
\begin{equation}  \label{eq:mu} 
 \mu(C_{0} \cup C_1) = (|C_0|\mu(C_0)+ |C_1|\mu(C_1))/ (|C_0|+|C_1|)  
\end{equation}
 \begin{tiny} $$\sigma^2(C_{0} \cup C_1) = \dfrac{|C_0|\sigma^2(C_0)+ |C_1|\sigma^2(C_1)+  \frac{|C_0||C_1|}{|C_0|+|C_1|}(\mu(C_0)-\mu(C_1))^2}{ |C_0|+|C_1|}$$ \end{tiny}



\subsection{Parameter-free RP-ALC algorithm} \label{sec:NoParaALC} 
The parameter-free RP-ALC algorithm is an extension of the parameter-free RP-SLC algorithm. However, we must use the ALC-distance, and handle insertions and removals of points(clusters) due to mergers. We use adapted definitions for feasible and taken edges of those in Section \ref{sec:NoPara}. There are two main differences for ALC: First, the sets $\m{S} \in \mf{S}$ get sparser with every merger, i.e., for every merger two clusters $C_0,C_1$  are replaced by a new single center $C'$. Every set $\m{S} \in \mf{S}$ with either $C_0 \in \m{S}$ or $C_1 \in \m{S}$ is replaced by $\m{S}':=\m{S}\setminus \{C_0,C_1\} \cup C'$.  

\begin{figure}[htp]
\centerline{\includegraphics[width=1\linewidth]{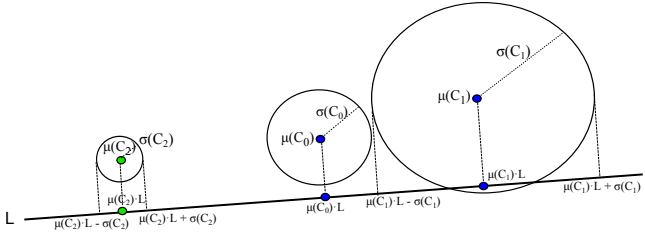}}

\caption{The figure illustrates that two clusters $C_0,C_1$ might have a larger ALC-distance (see definition \ref{eq:alc}) than two clusters $C_0$ and $C_2$ although the center $\mu(C_1)$ is closer to $\mu(C_0)$ than to $\mu(C_2)$. } 
\label{fig:avgLink}
\end{figure}

Second, it might be the case that the clusters $C_i,C_j$ to be merged do not have the cluster centers $\mu(C_i),\mu(C_j)$ with shortest distance $D(\mu(C_i),\mu(C_j))$ as also the variance must be accounted for (see Figure \ref{fig:avgLink}).

To deal with these two problems, we state conditions on the minimum frequency of edges within sets $\m{S}$ analogous to Condition \ref{eq:cond} but taking into account the cluster variances. 
 An inter-cluster edge $e=\{C_0,C_1\}$ that is frequent is not necessary feasible because of the variances, but it is potentially feasible. More formally, an edge $\{F,P\}$ is \emph{potentially feasible} if $\{F,P\} \in \m{S}$ for a fraction $c_f$ of all sets $\mf{S}$.  To ensure that cluster $C_1$ has minimum ALC-distance to $C_0$, it suffices to compare the ALC-distance between $C_0$ and $C_1$ with the ALC-distances between $C_0$ and all clusters $C_i$ for which it holds $D(C_0,C_i)=(\mu(C_i)-\mu(C_0))^2 \leq D(C_0,C_i)+\sigma(C_0)^2\leq (\mu(C_1)-\mu(C_0))^2 + \sigma(C_0)^2+\sigma(C_1)^2=D_{ALC}(C_0,C_1)$. One way to achieve that all these comparisons are made by the algorithm, i.e., that there is a set containing all these clusters $C_i,C_1,C_0$, is by requiring that there be a potentially feasible edge $\{C_0,C'\}$ with $(\mu(C')-\mu(C_0))^2\geq (\mu(C_1)-\mu(C_0))^2 + \sigma(C_1)^2$. In this case, all shorter edges $\{C_0,C_i\}$ measured using the ALC-distance must also be in any of the sets.


 An edge $e=\{F,P\} \in \m{S}$ is \emph{feasible} if it is potentially feasible and there is also a potentially feasible edge $e'=\{\hat{F},\hat{P}\} \in \m{S}$ such that $(\mu(F)-\mu(P))^2+\sigma(P)^2+\sigma(F)^2\leq (\mu(\hat{F})-\mu(\hat{P}))^2$.   An edge $\{T,P\}$ is taken if $\{T,P\} \in \m{S}$ for a fraction $c_f$ of all sets $\mf{S}$. We compute for each cluster $P$, i.e., represented by its center $P$, the potentially feasible edges to cluster (centers) $F \in \m{F}(P)$ and the taken edges to cluster (centers) $T \in \m{T}(P)$. We double $minPts$ until there is either a feasible or a taken edge for each point $P \in \m{P}$.
We deal with the sparsification by maintaining the following condition (analogous to Condition \ref{eq:cond} in Section \ref{sec:NoPara}):
\[ \begin{medsize}\begin{aligned}
 \label{eq:cond2} 
|\m{F}|>0 \wedge & \forall P \in \m{P}^{check}:  |\m{F}(P)|>0 \vee \\ & (|\m{T}(P)|>0 \wedge \max_{T \in \m{T}(P)} D_{ALC}(P,T)\geq 
\min_{\substack{Q \in \m{P}\\R \in \m{F}(Q)}} D_{ALC}(Q,R)) \nonumber
\end{aligned}  \end{medsize} \]

We can use algorithm RP-SLC with the modified definitions for feasible and taken edges to compute an ALC. Furthermore, after every merger the distances of the merged cluster to all other clusters co-occurring in a set $\m{S}$ must be considered. The analysis of the parameter-free RP-ALC algorithm is analogous to that of the parameter-free RP-SLC algorithm.


\section{Experimental Evaluation}\label{sec:exhaustivealgorithms}
For all experiments, we set $minPts=14$ and computed $|\mf{L}|=20\log N$ sequences of random projections for all benchmark datasets. We use both real and synthetic data sets from \cite{UCI10,Fin12}.

\medskip\noindent\textbf{HC preservation:}
To evaluate the quality of the resulting HC, we evaluate how \textit{isomorphic}
 the new dendrograms on the projected data are compared with those from the original data. We assess the
similarity between two dendrograms using the confusion matrix of the clusters formed when
`cutting' the dendrogram at different cluster resolutions \cite{fowlkes89}. Then, we take the average of cluster
affinities at all levels of the dendrogram.

Table \ref{table:res} reports the preservation of hierarchical clustering based on the above measure. 
Cluster preservation is consistently greater than $99\%$ across all datasets.

\begin{table}[h!]
\caption{The HCs (i.e. dendrograms) computed by the RP-SLC/ALC algorithms and traditional HC algorithms are almost identical.} 
\centering 
\footnotesize{
\begin{tabular}{l |l  |l } 
\hline\hline 
Dataset & SLC Preservation  & ALC Preserv. \\ [0.5ex] 
\hline 
Iris 		&  100\% &  100\% \\
Aggregation		& 100\% &  99.99\% \\
Glass &  100\% &  100\% \\
KDDCUP04Bio &  100\% & 100\%  \\
Covertype   &  99.98\% &  100\% \\
Pathbased &  100\% &  100\% \\
Synt.Control-TimeSeries &  100\% &  100\% \\
A1 &  100\% & 100\% \\
Dim512 & 100\% &  100\% \\
\hline 
\end{tabular}
}

\label{table:res}
\end{table}


\begin{figure}[!ht]
\centering
\includegraphics[width=\linewidth]{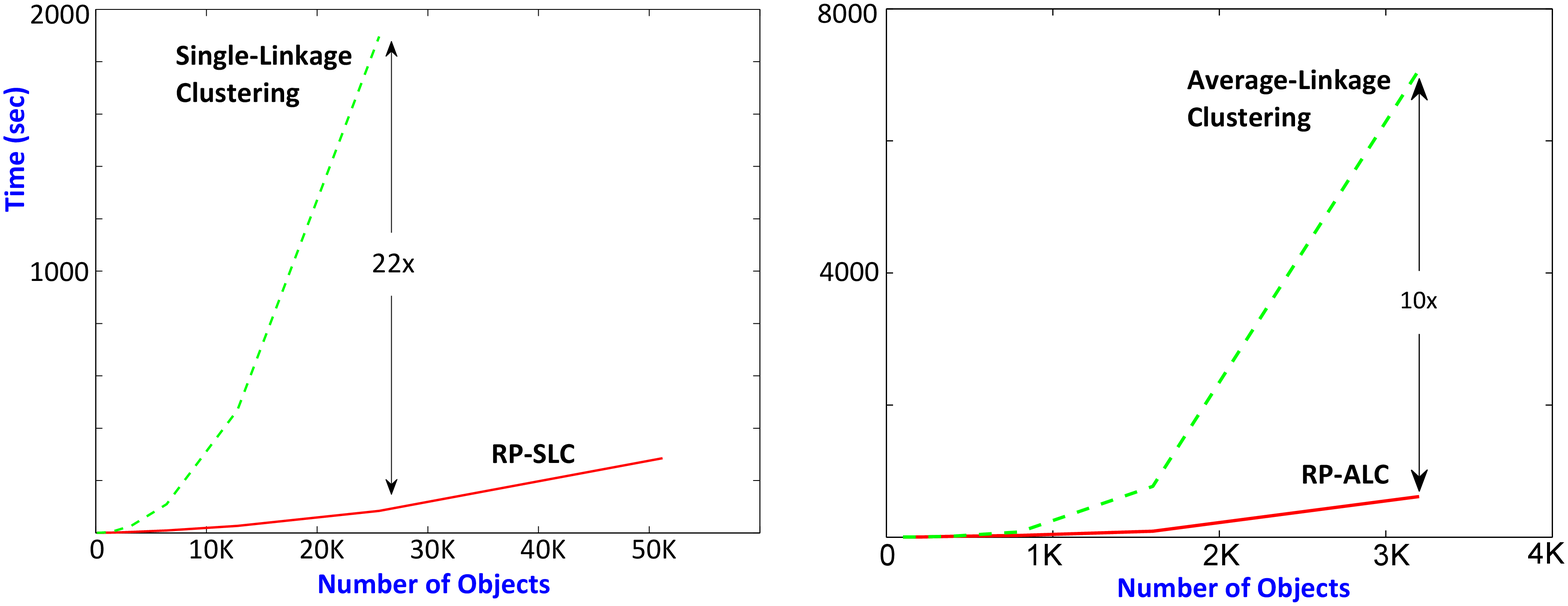}

\caption{Comparison of the RP-ALC/RP-SLC algorithm (dashed lines) and ALC/SLC algorithms (solid lines).}  
\label{fig:runTime}
\end{figure}

\medskip\noindent\textbf{Runtime improvement:}
We conduct a separate experiment to assess the improvement in runtime using synthetic datasets. Data clusters were created according to a Gaussian distribution in a 500-dimensional space. For SLC, our RP-SLC algorithm is more than 20x faster than traditional SLC algorithms, as shown in Figure \ref{fig:runTime}.  Our technique also scales significantly better in terms of space. For the standard ALC  (and the RP-ALC) algorithm, we maintained centroids and variances rather than the entire sets of cluster points (see Equation \ref{eq:mu}). Thus, the difference in runtime stems from the fact that RP-ALC computes and maintains fewer distances. 
The asymptotic gain of roughly a factor of $N$ in time-complexity by using our algorithms is apparent for both SLC and ALC.











\scriptsize{
\bibliographystyle{abbrv}
\bibliography{databases}

\begin{thebibliography}{10}

\bibitem{Fin12}
{Clustering datasets, Speech and Image Processing Unit, University of Eastern
  Finland}.
\newblock \url{http://cs.joensuu.fi/sipu/datasets/}.
\newblock Accessed: 4/5/2012.

\bibitem{Ran12}
{Creating points uniformly at random on N dim. sphere}.
\newblock \url{http://www.math.niu.edu/~rusin/known-math/96/sph.rand}.
\newblock Accessed: 4/5/2012.

\bibitem{UCI10}
{UCI Machine Learning Repository}.
\newblock \url{http://archive.ics.uci.edu/ml/datasets.html}.
\newblock Accessed: 4/5/2012.

\bibitem{Agr90}
P.~K. Agarwal, H.~Edelsbrunner, O.~Schwarzkopf, and E.~Welzl.
\newblock {Euclidean Minimum Spanning Trees and Bichromatic Closest Pairs}.
\newblock In {\em Computational Geometry}, 1990.

\bibitem{Day84}
W.~Day and H.~Edelsbrunner.
\newblock Efficient algorithms for agglomerative hierarchical clustering
  methods.
\newblock {\em Journal of Classification}, 1984.

\bibitem{Fern03}
X.~Z. Fern and C.~E. Brodley.
\newblock {Random Projection for High Dimensional Data Clustering: A Cluster
  Ensemble Approach}.
\newblock In {\em ICML}, 2003.

\bibitem{fowlkes89}
E.~B. Fowlkes and C.~L. Mallows.
\newblock A method for comparing two hierarchical clusterings.
\newblock {\em Journal of the American Statistical Association}, 78(383), 1983.

\bibitem{Gio00}
T.~H. Haveliwala, A.~Gionis, and P.~Indyk.
\newblock Scalable techniques for clustering the web.
\newblock In {\em WebDB (Informal Proceedings)}, 2000.

\bibitem{Mar10}
S.~Hoory, N.~Linial, and A.~Widgerson.
\newblock {Expander Graphs and their Applications}.
\newblock In {\em {Bulletin of the American Mathematical Society}}, 2006.

\bibitem{Kog07}
H.~Koga, T.~Ishibashi, and T.~Watanabe.
\newblock {Fast agglomerative hierarchical clustering algorithm using
  Locality-Sensitive Hashing}.
\newblock {\em Knowl. Inf. Syst.}, 2007.

\bibitem{Lei06}
B.~Leibe, K.~Mikolajczyk, and B.~Schiele.
\newblock {Efficient Clustering and Matching for Object Class Recognition}.
\newblock In {\em BMVC}, 2006.

\bibitem{Wil10}
W.~B. March, P.~Ram, and A.~G. Gray.
\newblock {Fast euclidean minimum spanning tree: algorithm, analysis, and
  applications}.
\newblock KDD, 2010.

\bibitem{Meh11}
K.~Mehlhorn, R.~Osbild, and M.~Sagraloff.
\newblock A general approach to the analysis of controlled perturbation
  algorithms.
\newblock {\em Computational Geometry}, 2011.

\bibitem{Mur84}
F.~Murtagh.
\newblock {Complexities of hierarchical clustering algorithms: state of the
  art}.
\newblock {\em Computational Statistics Quarterly}, 1984.

\bibitem{Sib73}
R.~Sibson.
\newblock {SLINK}: an optimally efficient algorithm for the single-link cluster
  method.
\newblock {\em The Computer Jo.}, 1973.

\bibitem{Vip05}
P.-N. Tan, M.~Steinbach, and V.~Kumar.
\newblock {\em {Introduction to Data Mining}}.
\newblock 2005.

\bibitem{Urr07}
T.~Urruty, C.~Djeraba, and D.~A. Simovici.
\newblock {Clustering by Random Projections}.
\newblock In {\em Ind. Conf. on Data Mining}, 2007.

\end{thebibliography}
}

\end{document}